\documentclass[runningheads]{llncs}\usepackage[]{graphicx}\usepackage[]{xcolor}
\makeatletter
\def\maxwidth{ %
  \ifdim\Gin@nat@width>\linewidth
    \linewidth
  \else
    \Gin@nat@width
  \fi
}
\makeatother

\definecolor{fgcolor}{rgb}{0.345, 0.345, 0.345}

\usepackage{framed}
\makeatletter
 {\par\unskip\endMakeFramed%
 \at@end@of@kframe}
\makeatother

\definecolor{shadecolor}{rgb}{.97, .97, .97}
\definecolor{messagecolor}{rgb}{0, 0, 0}
\definecolor{warningcolor}{rgb}{1, 0, 1}
\definecolor{errorcolor}{rgb}{1, 0, 0}
\newenvironment{knitrout}{}{} 

\usepackage{alltt}
\usepackage[T1]{fontenc}
\usepackage{graphicx}
\IfFileExists{upquote.sty}{\usepackage{upquote}}{}
\begin{document}
\title{All edges lead to Rome: analyzing Late Antiquity regime shifts using coin hoard data}
\titlerunning{All edges lead to Rome}
%
\author{Juan J. Merelo-Guervós\inst{1}\orcidID{0000-0002-1385-9741}}
\authorrunning{JJ Merelo}
%
\institute{Department of Computer Engineering, Automatics and Robotics and CITIC, University of Granada
\email{jmerelo@ugr.es}\\
}
\maketitle              
\begin{abstract}
Complex network analysis and other statistical techniques can be fruitfully applied to historical data to understand dynamics that brought the territory under study to a certain state. In this paper we use the FLAME database to understand when a drastic change of trade regime occurred in the Italian peninsula by looking at mint-hoard links that are evidence of a physical connection between them. By using changepoint analysis to pinpoint precise dates in the time series of these links, and then historical as well as network data to establish the sequence of network collapse, we find that the demise of the Western Roman Empire started at the beginning of the 5th century AD, yielding a more precise sequence of how and when it reached Italia.

\end{abstract}
Until the arrival of cliometrics \cite{cliometrics}, history was largely a qualitative affair that enabled historians and historiographers to create narratives based largely on historical sources with their corresponding bias and credibility. Modern historians, starting with Pirenne \cite{pirenne}, advocated for a {\em tout court}, or whole field, approach to history that included input from other sciences; using complex networks in this area is but a small added value to this sequence; however, given the chance to merge in a single model an extended time and territory span, it has been so far a fruitful approach.

The caesura between Antiquity and the Middle Ages, generally coincident with the fall of the Roman Empire in AD 476, is one such event. The deposition of the Western Roman Emperor marked the beginning of a new era where there was no longer a unified polity that spanned the whole Mediterranean basin and beyond. Or did it? That is, is it possible to establish a sequence of events that led to that foregone conclusion?

There are different ways of analyzing the Roman Empire using complex network analysis, but one that has proven fruitful so far is to use the FLAME database \cite{FLAME:bias}. This database, last updated in April 2026, contains geolocated coin hoard data, identifying groups of coins by their geolocated mint. This dataset is processed to attribute every coin group and mint to a territory, which is designated, in most cases, by its traditional name (which can include several countries, like the Iberian Peninsula). Coin hoards have a date range. We attribute a proportion of the total to every year in the date range to create a dataset that has a mint, a hoard, a year and a {\em link rate}; these compose a time series of trade-like links.

We have so far used this data to analyze regime changes in the global network (in its 2024 state) \cite{changepoint.statphys} and the network centered on the Iberian Peninsula \cite{complexities2020012}. In these papers we apply different changepoint analysis methods to find changes of regime in the network of mint-hoard links, and then analyze the resulting network to check what the regime change has brought. What the analysis revealed is that regime change preceded the fall by almost a century, and that it was provoked by node drops (Britannia, \cite{complexities2020012}) creating a cascade of failures, and edge drops occurring elsewhere \cite{changepoint.statphys} provoking a modularization that might eventually result in a network split.

In this paper, we will look at the data from the center of the network: Italia, the generic and historic name for the Italian Peninsula, which hosted the capital of the Western Roman Empire until its fall. By examining regime changes from this point of view, we will be able to examine more closely which network edges were essential and, by vanishing, precipitated its downfall.

\begin{knitrout}
\definecolor{shadecolor}{rgb}{0.969, 0.969, 0.969}\color{fgcolor}\begin{figure}[h!tb]
\includegraphics[width=0.5\linewidth]{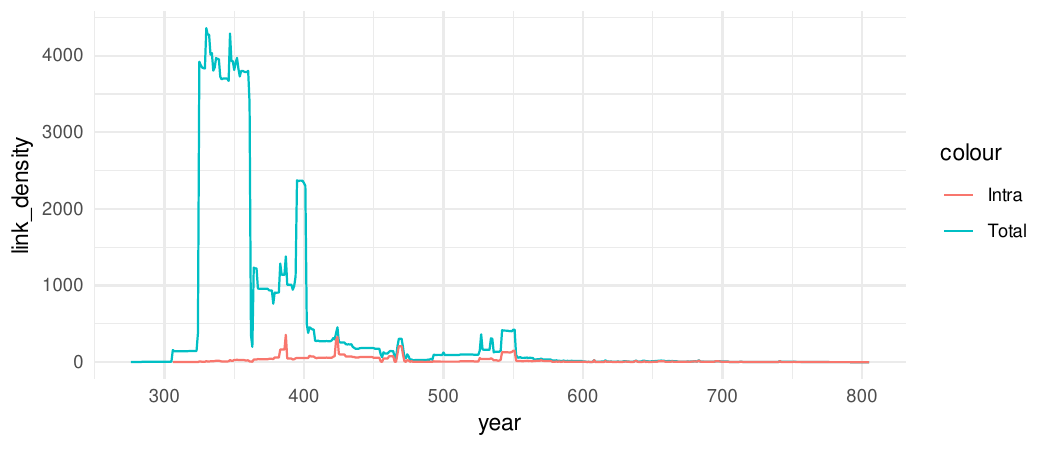} 
\includegraphics[width=0.5\linewidth]{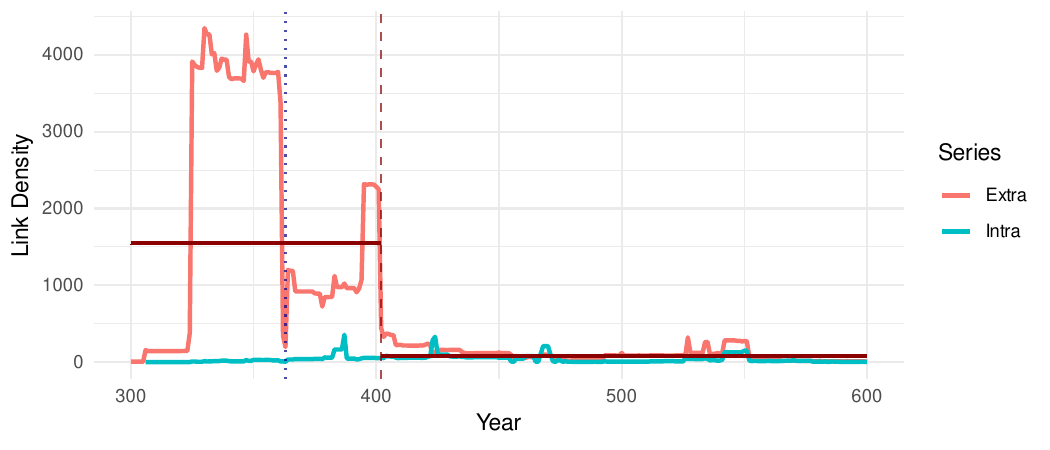} \caption[Time series of link density for all links with Italia, as well as internal links]{Time series of link density for all links with Italia, as well as internal links. (Right) Changepoints detected over the intra and extra trade link series, dashed (trade links only), dotted (with entropy). Averages pre and post the first changepoint are shown in dark red.}\label{fig:wivace26.series}
\end{figure}

\end{knitrout}

Figure \ref{fig:wivace26.series} shows the internal and general time series of link density; it spans the whole dataset. As we see, there are profound changes before the purported fall of the Empire.
In order to analyze the changepoint, we will create three time series: intra-Italia trade density, extra-Italia trade density (obtained by subtracting the intra-Italia density from the total), and per-year entropy, an interesting indicator of the variability of trade sources, as well as the position of the node in the network.

We will first analyze the multi-stream series that includes only trade density using {\sf e.divisive} from {\sf ecp} \cite{ecp-article}, and a minimum size of 100 years. It generates several changepoints; the most important, showing the largest shift in average trade density before and after, falls in the year 402. After including the entropy time series and re-scaling, the year for the change becomes 363 AD (represented in the right-hand side of Figure \ref{fig:wivace26.series}). We observe that the average exterior trade density falls from 1549.1 to 77.45 after the first changepoint, which occurs at 402.

It is interesting to see also how the entropy behaves by itself, since it moves the changepoint to a later date. Using the same method, the changepoint for the entropy time series is detected in 376.

\begin{knitrout}
\definecolor{shadecolor}{rgb}{0.969, 0.969, 0.969}\color{fgcolor}\begin{figure}[h!tb]
\includegraphics[width=\maxwidth]{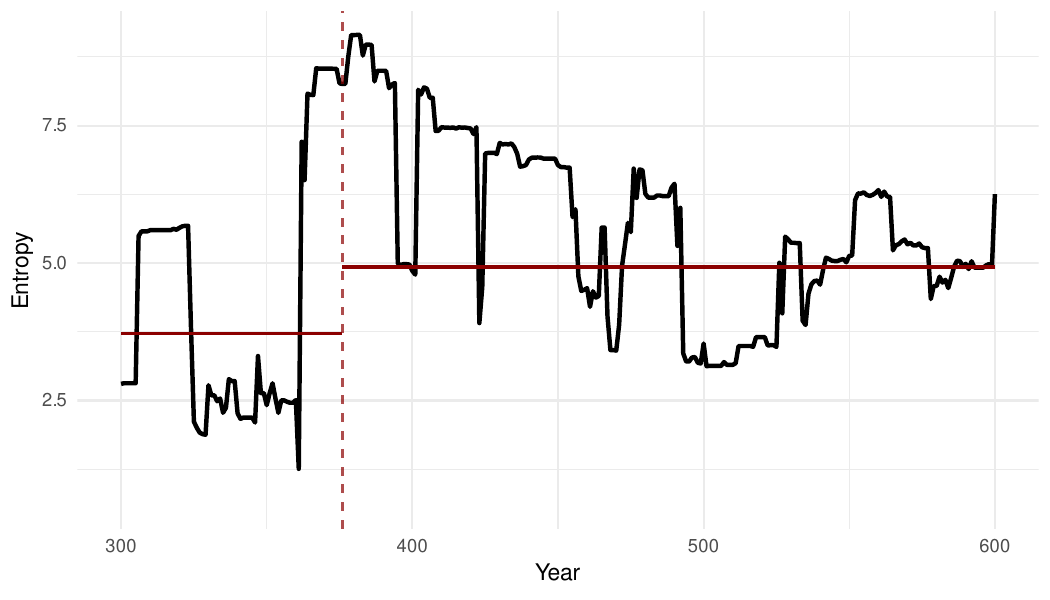} \caption[Entropy of the Italia ego network]{Entropy of the Italia ego network. The dashed line shows the changepoint detected}\label{fig:postwivace26.entropy_plot}
\end{figure}

\end{knitrout}

Figure \ref{fig:postwivace26.entropy_plot} chars entropy, including the changepoint, which occurs years before the main changepoint involving the two time series. And entropy actually goes up: the value before that changepoint is 3.72; it increases after that date to 4.93. So the number of trade partners and their variability actually start to increase decades before the already commented change of regime occurs.

In terms of the network, what does this change imply?
\begin{knitrout}
\definecolor{shadecolor}{rgb}{0.969, 0.969, 0.969}\color{fgcolor}\begin{figure}[h!tb]
\includegraphics[width=0.5\linewidth]{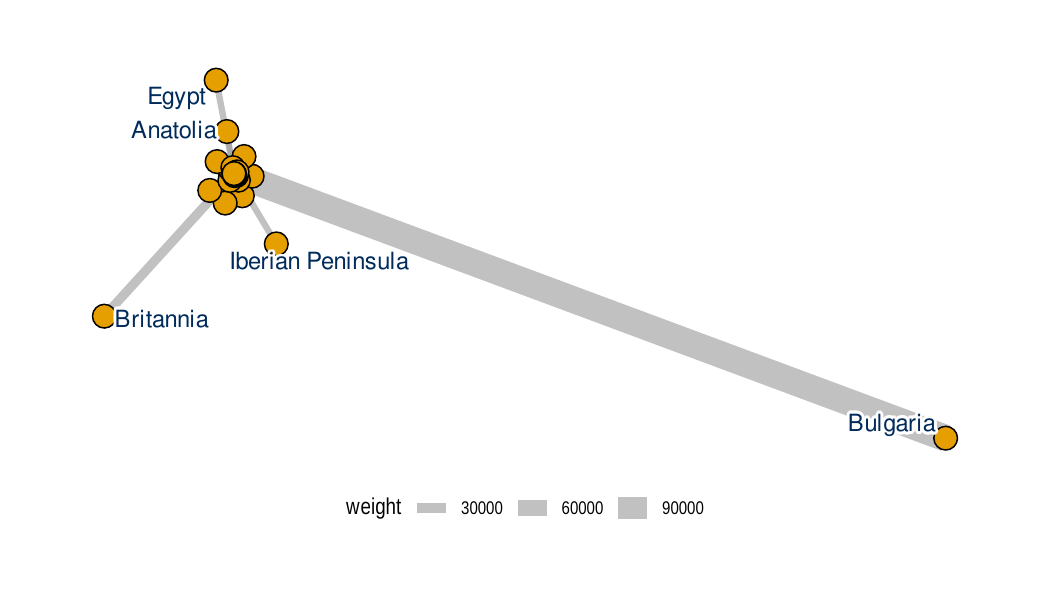} 
\includegraphics[width=0.5\linewidth]{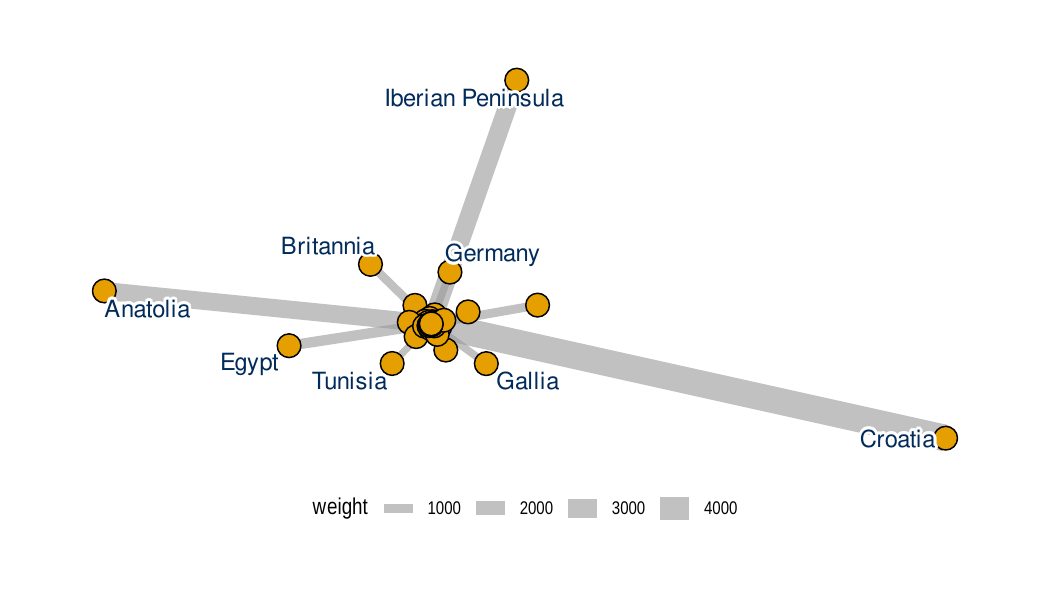} \caption[Representation of the Italia ego network, before the changepoint (left) and after (right)]{Representation of the Italia ego network, before the changepoint (left) and after (right). Edges are scaled according to link density value. Scales shown at bottom.}\label{fig:wivace26.graphs}
\end{figure}

\end{knitrout}

Figure \ref{fig:wivace26.graphs} shows again the different scales before and after the changepoint, but the connection from Italia to the territory of Bulgaria, whose value is much higher than that of any other connection before 402, is much more evident. Bulgaria contained, at that time, the territory called Thrace, which was the land connection to Constantinople and belonged to the Eastern Roman Empire. There are also more nodes in the ego network before than after: 51 vs. 47. This ``loss'' of Thrace matches the analysis in \cite{changepoint.statphys} over the global network: breaking the Danube as a network edge provoked the creation of massive structural holes covering Central Europe and the Balkans.

\begin{knitrout}
\definecolor{shadecolor}{rgb}{0.969, 0.969, 0.969}\color{fgcolor}\begin{figure}
\includegraphics[width=\maxwidth]{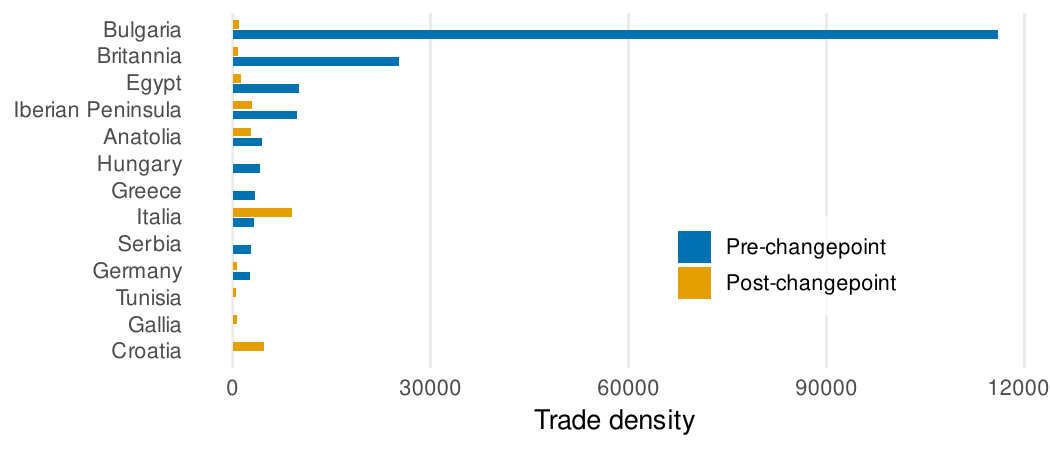} \caption[Change in total link density pre and post changepoint]{Change in total link density pre and post changepoint}\label{fig:wivace26.chart.trade}
\end{figure}

\end{knitrout}

We drill down into the reconfiguration of the network in Figure \ref{fig:wivace26.chart.trade} that shows the union of the top ten trading partners before and after the changepoint. We also see the importance of Britannia, which was already examined in \cite{complexities2020012} for its importance in the regime shift in the Iberian Peninsula. Since Italia was the ``center'' of the network, the network collapse initiated in these two places eventually affected Italia at a relatively early date. Internal trade is the only one that retains any significant weight after the changepoint, with other places like Croatia (possibly an alternative point of contact with the still-functioning Eastern Roman Empire), Anatolia and Egypt having also some vestigial trade at a very small fraction of the original. The fact that Bulgaria, which is the land bridge to the capital of the Eastern Roman Empire, Constantinople, has less trade than Anatolia itself is witness to the fact that the land route through the Military Road that followed the Danube was no longer operative.

We have just shown how complex network analysis, but above all, vision, can be applied to better understand historical dynamics, once they are combined with statistical techniques such as changepoint analysis, that detects changes of regime in single or synchronized time series. It allows us to create a narrative for the fall of the Roman Empire backed by numerical analysis: the defeat in the Battle of Adrianople, followed by the first withdrawal of Roman troops from Britannia at the end of the 4th century \cite{EmergingPowersinEurasianComparison2001100} and its definitive withdrawal a few years later, precipitated a collapse in the network that eventually left Italia divided and split from the rest of the European countries and the Mediterranean basin.

\begin{credits}
\subsubsection{\ackname} This work is supported by the Ministerio espa\~{n}ol de Econom\'{\i}a y
Competitividad (Spanish Ministry of Competitivity and Economy) under project PID2023-147409NB-C21.

Source for this paper can be downloaded from \url{https://github.com/JJ/medieval-trade}.

\end{credits}

\bibliographystyle{splncs04}
\bibliography{trade-early-medieval,ours,change-point,history}

\end{document}